\documentclass[superscriptaddress, reprint, amsmath, amssymb, aps, pra, floatfix]{revtex4-2}

\usepackage[utf8]{inputenc}
\usepackage{graphicx}
\usepackage{dcolumn}
\usepackage{bm}
\usepackage[normalem]{ulem} 
\usepackage{mathrsfs}
\usepackage{physics}
\usepackage{xcolor}
\usepackage{amsmath, amssymb, mathtools, amsthm}
\usepackage[colorlinks=true]{hyperref}
\usepackage{comment} 
\usepackage{orcidlink}

\usepackage{xr}

\graphicspath{{Figures/}} 

\definecolor{ve}{RGB}{0,161,22}

\begin{abstract}
Black holes permit matter to cross their event horizon in only one direction. We show that interacting bosons in optical lattices with asymmetric barrier exhibit an analogous phenomenon, creating unidirectional quantum transport without external driving or dissipation. This directionality emerges purely from many-body interactions, which cause asymmetric projection of the initial state onto transport-enabled or transport-forbidden sectors. The resulting dynamics create an effective one-way boundary in Hilbert space, forming a quantum analog of a black-hole event horizon. Our results establish interactions as a fundamentally new route to directional transport, enabling coherent rectification in atomtronic circuits by the use of intrinsic properties of the system only.
\end{abstract}

\begin{document}

\author{Elvira Bilokon\orcidlink{0009-0007-8296-2906}}
\email{ebilokon@tulane.edu}
\affiliation{Department of Physics and Engineering Physics, Tulane University, New Orleans, Louisiana 70118, United States}
\affiliation{Akhiezer Institute for Theoretical Physics, NSC KIPT, Akademichna 1, 61108 Kharkiv, Ukraine}

\author{Valeriia Bilokon\orcidlink{0009-0001-1891-0171}}
\email{vbilokon@tulane.edu}
\affiliation{Department of Physics and Engineering Physics, Tulane University, New Orleans, Louisiana 70118, United States}
\affiliation{Akhiezer Institute for Theoretical Physics, NSC KIPT, Akademichna 1, 61108 Kharkiv, Ukraine}

\author{Frank Gro\ss mann}
\affiliation{Institut f\"ur Theoretische Physik, Technische Universit\"at Dresden, D-01062 Dresden, Germany}

\author{Jason R. Williams\orcidlink{0000-0002-3798-4424}}
\affiliation{Jet Propulsion Laboratory, California Institute of Technology, Pasadena, CA 91109, USA}

\author{Denys I. Bondar \orcidlink{0000-0002-3626-4804}}
\email{dbondar@tulane.edu}
\affiliation{Department of Physics and Engineering Physics, Tulane University, New Orleans, Louisiana 70118, United States}

\title{Hilbert Space Black Hole Analog: Unidirectional Transport without Driving}

\maketitle

\section{Introduction}
Astrophysical black holes permit matter to flow in one direction only---an extreme manifestation of broken transport symmetry, where particles cross an event horizon but cannot return.
Black hole analogs have been created in systems ranging from acoustic horizons in Bose-Einstein condensates~\cite{Garay2000, Balbinot2008, Carusotto2008, Macher2009, Lahav2010, Steinhauer2016, Caio2022, Caio2023} to optical event horizons in nonlinear media~\cite{Philbin2008, Demircan2011, Elazar2012, Finazzi2013, Drori2019}, water wave analogs~\cite{Rousseaux2008, Weinfurther2011, Euve2016, Euve2020, Rozenman2024}, superconductor and superfluid interfaces~\cite{Manikandan2017, Manikandan2018}, and lattice models in superconducting qubits~\cite{Shi2023} where effective metric descriptions mimic gravitational spacetime curvature. Recent work has also explored black-hole analogue dynamics in controllable lattice models, including simulations of curved-spacetime Dirac fields and scrambling dynamics in engineered spin chains~\cite{Liu2025SciChina, Liu2025AdvSci}.
While gravity provides the most dramatic realization of such irreversibility, the broader question of how transport acquires a preferred direction extends across physics, from electronic devices to quantum matter.

Directional transport is conventionally achieved through ratchet mechanisms, nonlinear response, or time-dependent driving~\cite{Hanggi2009, Rousselet1994, Astumian2002, Reimann2002, Linke1999}. In quantum systems, directionality has been demonstrated using dissipation~\cite{Mennerat1999}, coupling to external reservoirs~\cite{Sanchez2004}, periodic modulation~\cite{Lignier2007, Denisov2014}, or topological mechanisms~\cite{Lohse2016, Nakajima2016}. These approaches—whether classical or quantum—predominantly operate at the single-particle level, relying on either explicit time-reversal symmetry breaking through dissipation or engineered time-dependent potentials~\cite{Salger2009, Denisov2014, Creffield2011}. The emergence of directional transport in closed, undriven quantum systems has remained an outstanding challenge. This raises a natural question: is there a fundamentally different route to directionality, one that does not rely on driving, dissipation, or reservoirs?

Interactions fundamentally alter this picture. While many-body effects are often associated with transport suppression through self-trapping~\cite{Smerzi1997, Albiez2005, Anker2005, Mistakidis2024, Alana2025} or many-body localization~\cite{Basko2006, Nandkishore2015, Schreiber2015, Sierant2025}, interactions can also enable qualitatively new transport phenomena~\cite{Eckstein2009, Bilokon2025}. In systems combining spatial asymmetry with strong correlations, interactions restructure the accessible eigenstate manifold, potentially creating transport pathways unavailable to single particles. Previous studies of interaction-assisted tunneling~\cite{Morales2008} and correlated transport in optical lattices~\cite{Ronzheimer2013, Haller2010} have hinted at this possibility, but robust unidirectional transport in static, closed quantum systems has not been demonstrated.

In the autonomous, single particle case, tunneling probabilities are known to be symmetric (i.\ e.\ transmission as well as reflection probabilities at a fixed energy are the same from left to right and from right to left) even for asymmetric potentials, as  proven~\cite{Landau1981, Tann, Shegelski2020}.
In this study, we show that interacting bosons in a one-dimensional optical lattice with a static asymmetric barrier can exhibit one-way quantum transport, however: particles tunnel readily when incident from one direction while remaining trapped from the opposite direction. This behavior emerges from the interplay between barrier asymmetry, many-body interactions, {and initial localization}---without time-dependent driving, dissipation, or coupling to external reservoirs. Through eigenstate analysis, we reveal that the asymmetric potential selectively projects initial states onto manifolds with fundamentally different transport properties, creating a dynamical horizon in Hilbert space---a quantum analog of the black-hole event horizon. Unlike previous black hole analogs that simulate metric effects in physical space, this Hilbert space horizon arises from interaction-driven eigenstate selection rather than spacetime geometry.  Our results have direct implications for atomtronic circuit elements~\cite{Amico2021, Amico2022}, quantum state preparation~\cite{Gross2017, Kaufman2021} in optical lattices, and the design of interaction-controlled quantum devices that exploit many-body effects rather than external driving or dissipation.

\section{Model and Methods}
\begin{figure}
    \centering
    \includegraphics[width=\linewidth]{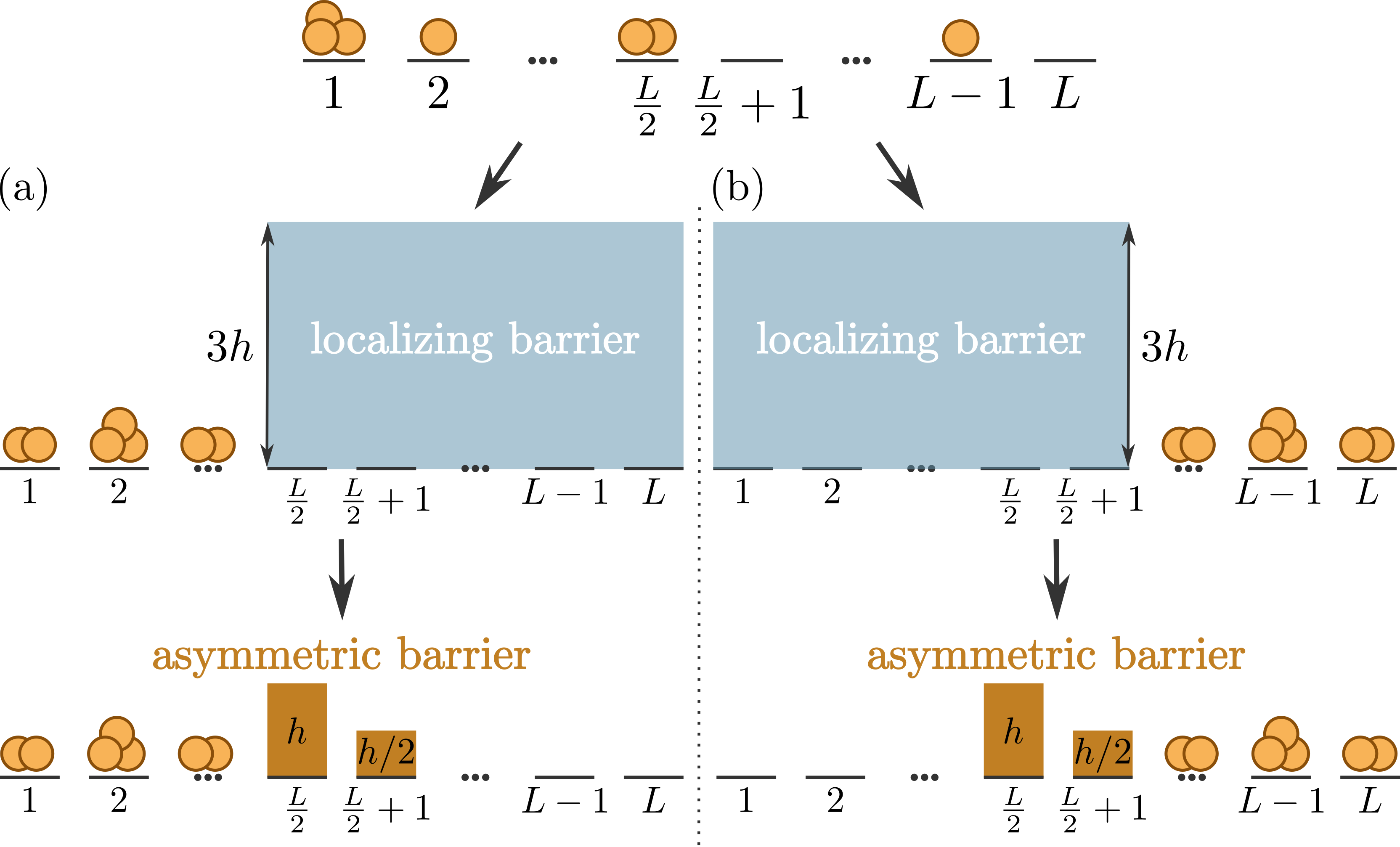}
    \caption{Initial state preparation protocol. Starting from a random distribution of $N$ bosons across the $L$-site lattice (top), a strong localizing potential ($3h$, shown in blue) is applied to confine bosons either (a) to the leftmost $L/2-1$ sites or (b) to the rightmost $L/2-1$ sites. The system is then evolved under the fixed asymmetric triangular barrier with heights $h$ and $h/2$ at sites $L/2$ and $L/2+1$, respectively. Particles initialized from the left encounter the vertical side first, while particles initialized from the right encounter the angled side first.}
    \label{fig:model}
\end{figure}

In our theoretical framework, we investigate the quantum dynamics of ultracold bosons confined to a one-dimensional optical lattice of $L$ sites, described by the Bose-Hubbard Hamiltonian
\begin{eqnarray}\label{eq:BHM}
    \hat{\mathcal{H}} &=& -J \sum_{j=1}^{L-1} \left( \hat{b}_{j}^{\dagger} \hat{b}^{}_{j+1} + \hat{b}_{j+1}^{\dagger} \hat{b}^{}_{j} \right) \\
    \nonumber
    &&+ \frac{U}{2}\sum_{j=1}^L \hat{n}_j(\hat{n}_j - 1) + \sum_{j=1}^L V^{\rm ex}_j\hat{n}_{j} \,.
\end{eqnarray}
Here, the operator $\hat{b}_{j}^{\dagger}$ ($\hat{b}^{}_{j}$) creates (annihilates) a boson at lattice site~$j$, satisfying the canonical commutation relations $[\hat{b}_i, \hat{b}_j^{\dagger}] = \delta_{ij}$. The number operator is defined as $\hat{n}_{j}=\hat{b}_{j}^{\dagger}\hat{b}^{}_{j}$. The first term in Eq.~\eqref{eq:BHM} represents the kinetic energy arising from the tunneling of bosons between neighboring sites with amplitude $J$. {Throughout, we set $J=1$ as the unit of energy, expressing all quantities in units of $J$}. The second term describes the on-site repulsive interaction with strength $U$, while the final term incorporates a site-dependent external potential $V^{\rm ex}_{j}$, which can be engineered to create specific trapping geometries. In our study, we consider the following barrier configuration:
\begin{align}
V^{\rm ex}_{j} &= h\delta_{j,L/2} + \frac h2 \delta_{j,L/2+1},
\label{eq:ext_pot} 
\end{align}
where $h$ denotes the barrier height and $\delta_{j,k}$ is the Kronecker delta function (see Fig.~\ref{fig:model}). We mention explicitly that in the above Hamiltonian, we use open boundary conditions.

{To prepare the initial quantum state, we employ a two-step protocol that ensures the bosons are initially localized away from the barrier region [see Fig.~\ref{fig:model}]. First, we introduce a localizing barrier configuration where all sites experience a strong confining potential of height $3h$, except for the sites on one side of the asymmetric barrier which remain at zero potential. The precise value $3h$ is chosen to be sufficiently large compared to the tunneling scale $J$ and the barrier height $h$, ensuring negligible occupation of the high-potential region. To study transport from the left~[see Fig.~\ref{fig:model}(a)], the localizing barrier confines bosons to sites 1 through $L/2-1$; to study transport from the right~[see Fig.~\ref{fig:model}(b)], a mirrored localizing barrier confines bosons to sites $L/2+2$ through $L$. We then obtain the ground state of the system under the respective localizing barrier configuration. In the second step, we perform a quantum quench by instantaneously switching from the localizing barrier to the fixed asymmetric triangular barrier given by Eq.~\eqref{eq:ext_pot}. This preparation protocol allows us to study transport through a single, fixed barrier Hamiltonian for both incident directions, directly demonstrating the emergence of unidirectional transport.}

The key observable for characterizing directional transport is the population imbalance
\begin{equation}
    \Delta n(t) = n_{\rm after}^{\rm (L)}(t) - n_{\rm after}^{\rm (R)}(t),
\end{equation}
{where $n_{\rm after}^{\rm (L)}(t) = \sum_{j=L/2+2}^L\langle\hat{n}_j\rangle$ and $n_{\rm after}^{\rm (R)}(t) = \sum_{j=1}^{L/2-1}\langle\hat{n}_j\rangle$ measure the particle number in the post-barrier region for configurations depicted in Fig.~\ref{fig:model}(a) and Fig.~\ref{fig:model}(b), respectively}.

The quantum dynamics are computed using the QuSpin package~\cite{Weinberg2017, Weinberg2019}, which provides exact diagonalization and time evolution capabilities for quantum many-body systems. The exact diagonalization approach allows us to fully resolve the quantum many-body dynamics without approximation, capturing all correlation effects and entanglement generation during the tunneling process.

\section{Emergence of Unidirectional Transport}
We start our discussion by characterizing the interaction-driven directional transport in the asymmetric triangular barrier system. For a lattice of $L=6$ sites with $N=4$ bosons, we fix $h=10J$ and systematically vary the interaction strength $U$ to map out the parameter regime where unidirectional tunneling emerges. 

This system size is chosen to balance computational tractability with the emergence of many-body effects: the system is small enough to permit exact diagonalization of the full Hilbert space, yet large enough to capture interaction-driven collective dynamics. The filling fraction $N/L=2/3$ ensures an average occupation of approximately 2 bosons per site in the pre-barrier region, thereby avoiding self-trapping effects that would suppress tunneling at higher occupations. While the choice of $h$ is made for concreteness, the unidirectional transport phenomena we describe are not restricted to this particular barrier height and persist across a range of $h$ values.

\begin{figure}
    \centering
    \includegraphics[width=\linewidth]{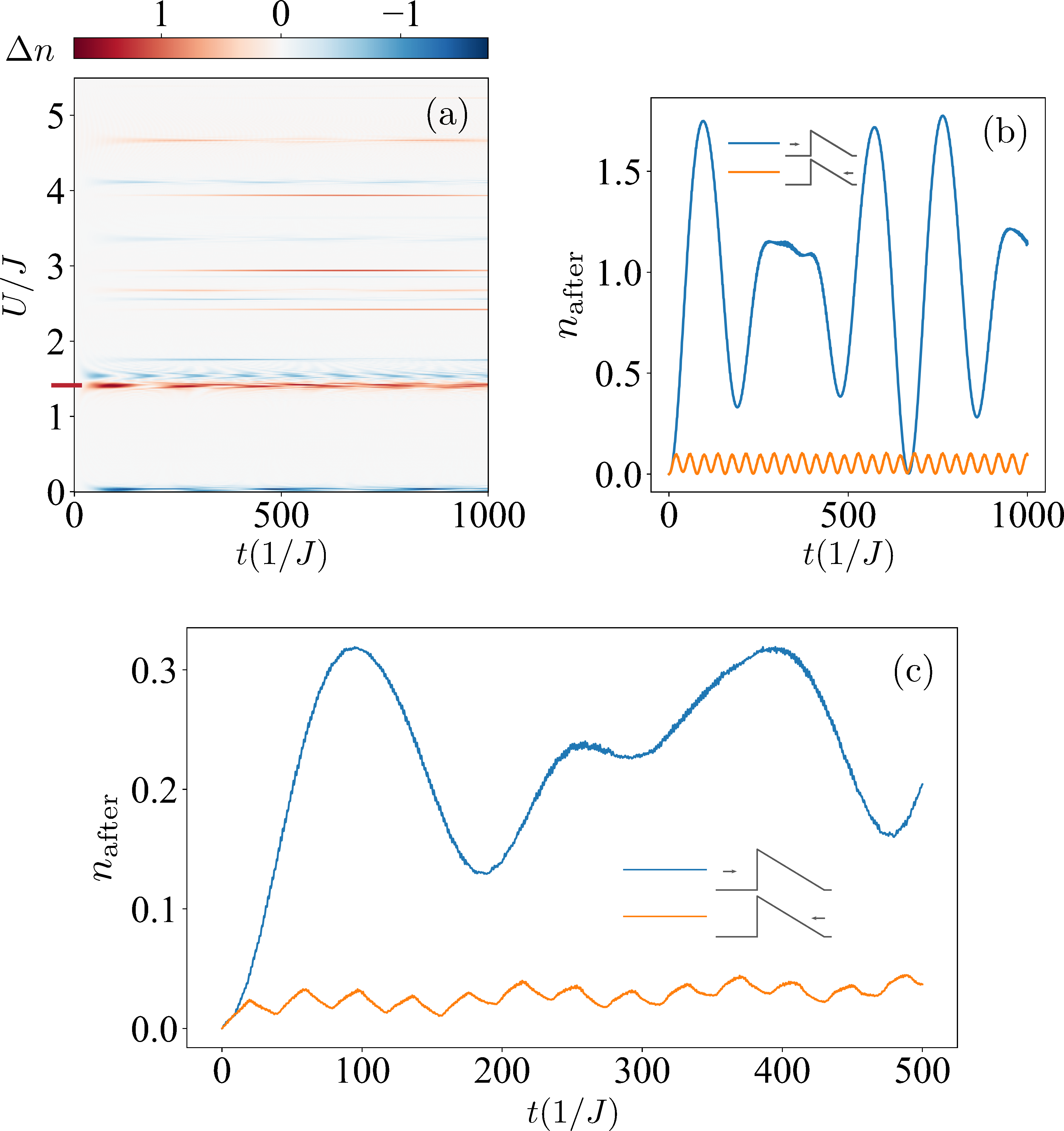}
    \caption{Directional transport in the asymmetric Bose-Hubbard model. (a) Heatmap showing the population imbalance  $\Delta n$ as a function of interaction strength $U$ and time $t$ for $N=4$ bosons, $L=6$ sites, barrier height $h=10 J$, and open boundary conditions. Red (blue) regions indicate preferential tunneling from the vertical (angled) side. The strongest directional transport occurs at $U \approx 1.42J $ (dark red band). Time evolution of $n_{\rm after}$ for $U=1.42J$ in case of (b) Fock state and (c) coherent state initial conditions. Blue curve: tunneling from vertical side, indicating collective many-body transport. Orange curve: tunneling from angled side remains suppressed.
}
    \label{fig:pop_imbalance_L6}
\end{figure}

Figure~\ref{fig:pop_imbalance_L6}(a) demonstrates the population imbalance $\Delta n$ as a function of time $t$ and interaction strength $U$. The heatmap reveals interaction windows with strong directional transport, where particles tunnel readily from one side of the barrier but remain trapped from the other side. The strongest directional asymmetry occurs at $U\approx1.42 J$, where positive $\Delta n$ indicates particles tunnel from the vertical side but not from the angled side. We focus our analysis on this regime, which exhibits robust directional transport over an extended parameter range ($\Delta U\sim 0.05 J$, corresponding to $3.5\%$ relative uncertainty in $U$). Narrower resonance-like features visible at higher $U$ values are excluded from detailed analysis due to their sensitivity to parameter variations. The required precision $\delta U / U = 3.5\%$ is readily achievable via Feshbach resonance control within current experimental capabilities~\cite{Marte2002, Borkowski2023}. 
Figure~\ref{fig:pop_imbalance_L6}(b) demonstrates the directional asymmetry at the interaction strength $U=1.42 J$: {particles tunnel from left with large-amplitude oscillations, while tunneling from right exhibits nearly complete suppression, with after-barrier} population remaining below 0.1 particles throughout the evolution. This dramatic contrast exemplifies the unidirectional transport regime where the interplay between asymmetric barriers and interactions creates effective one-way transport. {To emphasize the robustness of the effect, we numerically studied the quantum dynamics also for coherent state initial conditions $|n_1,\dots,n_L\rangle=\Pi_{j=1}^L{\rm exp}\left(-n_j^2/2+\sqrt{n_j}\hat b_j^\dagger\right)|{\bf 0}\rangle$ where $n_j$ are nonzero only on the left ($n_1,\dots,n_{L/2-1}$) or only on the right ($n_{L/2+2},\dots,n_{L}$) of the fixed asymmetric barrier, corresponding to left- and right-initialized configurations, respectively, i.e., for an initial wavefunction that is a direct product of a Poisson distribution over number (Fock) states in each well on the side of the barrier where the coherent state is prepared.}
This initial state is the closest quantum analog to a classical phase-space point as well as an excellent approximation to Bose-Einstein condensates prepared on single particle modes \cite{Lieb2007}. The quantum dynamics was solved for by using a fully variational approach based on a multi-configuration ansatz in terms of Glauber coherent states \cite{Werther2020} (for an implementation of the same initial states but using generalized coherent states as basis functions, see \cite{Qiao2025}). 
{For the same model parameters as the ones used in Fig.~\ref{fig:pop_imbalance_L6} and for $n_1=n_2=2, n_3,\dots,n_6=0$ (left-localized initial state) or $n_5=n_6=2, n_1,\dots,n_4=0$ (right-localized initial state), we find qualitatively similar behavior, although the amplitude of the oscillation in the left-initialized case is suppressed, as depicted in Fig.~\ref{fig:pop_imbalance_L6}(c). The stark difference between the tunneling from the left and from the right is still present, however.}


\section{Spectral Origin of Transport Windows}
\label{sec:transport_windows}
The heatmap in Fig.~\ref{fig:pop_imbalance_L6}(a) reveals a rich structure of fine stripes alongside the broad directional transport window at $U\approx 1.42J$. To systematically characterize these features and understand their origin, we introduce the long-time average of $n_{\rm after}^{(k)}(t)$
\begin{equation}\label{eq:long_time_avg}
    W^{(k)}(U) = \sum_i |\langle\phi_i(U)|\psi^{(k)}(0)\rangle|^2 \langle\phi_i(U)|n^{(k)}_\mathrm{after}|\phi_i(U)\rangle,
\end{equation}
where $k\in \{{\rm L}, {\rm R} \}$ labels the two initial configurations, $\ket{\psi^{(k)}(0)}$ represent the initial state, and $|\phi_i(U)\rangle$ denotes the eigenstate of the post-quench Hamiltonian~\eqref{eq:BHM} with interaction strength $U$. This quantity captures the stationary contribution to post-barrier population. Namely, a large $W^{(k)}(U)$ indicates that the initial state projects substantially onto eigenstates with significant post-barrier occupation, enabling transport, while $W^{(k)}(U) \approx 0$ signals projection onto transport-forbidden eigenstates.
 
\begin{figure}
    \centering
    \includegraphics[width=\linewidth]{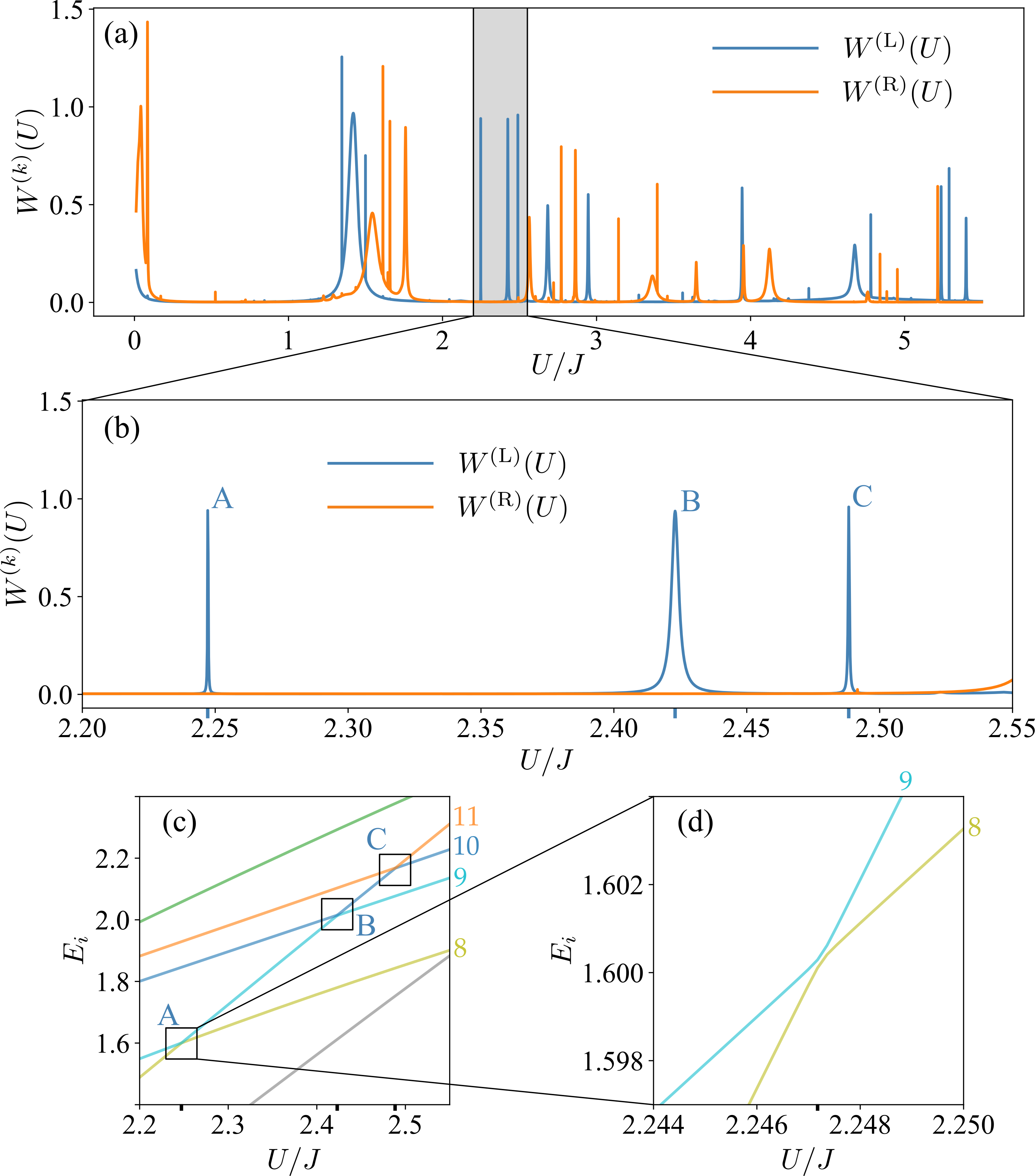}
    \caption{{(a)~Long-time average post-barrier population $W^{(k)}(U)$ [Eq.~\eqref{eq:long_time_avg}] for initial states with bosons localized from left (blue) and from right (orange). (b)~Zoomed view of $W^{(k)}(U)$ over the interval $U \in [2.20, 2.50]J$, showing three isolated peaks in $W^{(\mathrm{L})}(U)$ while $W^{(\mathrm{R})}(U)$ remains suppressed. (c)~Eigenenergies of Hamiltonian~\eqref{eq:BHM} in the same interaction window, revealing that each peak in $W^{(\mathrm{L})}(U)$ coincides with an avoided crossing. (d)~Detailed view of the avoided crossings near $U \approx 2.247J$, where the energy levels diverge rapidly producing a narrow transport window.}}
    \label{fig:transport_windows}
\end{figure}

Figure~\ref{fig:transport_windows}(a) shows $W^{\rm (L)}(U)$ and $W^{\rm (R)}(U)$ over the same interaction range used in Fig.~\ref{fig:pop_imbalance_L6}(a). Both configurations exhibit a rich, complex series of sharp resonant peaks distributed across the full interaction range. The peaks vary considerably in width and height: the widths correlate directly with the widths of the directional transport stripes visible in Fig.~\ref{fig:pop_imbalance_L6}(a), while the heights correlate with the color saturation in the same figure. We emphasize that the positions of the narrow peaks are highly sensitive to the sampling of $U$. In Fig.~\ref{fig:transport_windows}(a), the interaction interval $U \in [0, 5.5]J$ is sampled using 30001 points. At coarser resolution, narrow peaks can be missed entirely or appear merged, which is why the fine-grained sweep is necessary to resolve the resonance structure.

To understand the origin and varying widths of these peaks, Fig.~\ref{fig:transport_windows}(b) shows a zoomed view of $W^{(k)}(U)$ over the representative interval $U \in [2.20, 2.50]J$, where three isolated peaks of $W^{\rm (L)}(U)$  are clearly visible while $W^{\rm (R)}(U)$ remains suppressed. Figure~\ref{fig:transport_windows}(c) shows the corresponding eigenenergies of Hamiltonian~\eqref{eq:BHM} in the same interaction window. A direct comparison reveals that each peak in $W^{\rm (L)}(U)$ occurs at the interaction strength where the many-body energy levels approach one another and undergo avoided crossings. Near such level rearrangements, the eigenstate composition changes rapidly with $U$, which can strongly enhance the overlap of the initial state $\ket{\psi^{(k)}(0)}$ with eigenstates $\ket{\phi_i}$ that have nonzero post-barrier occupation. 

\begin{figure}
    \centering
    \includegraphics[width=\linewidth]{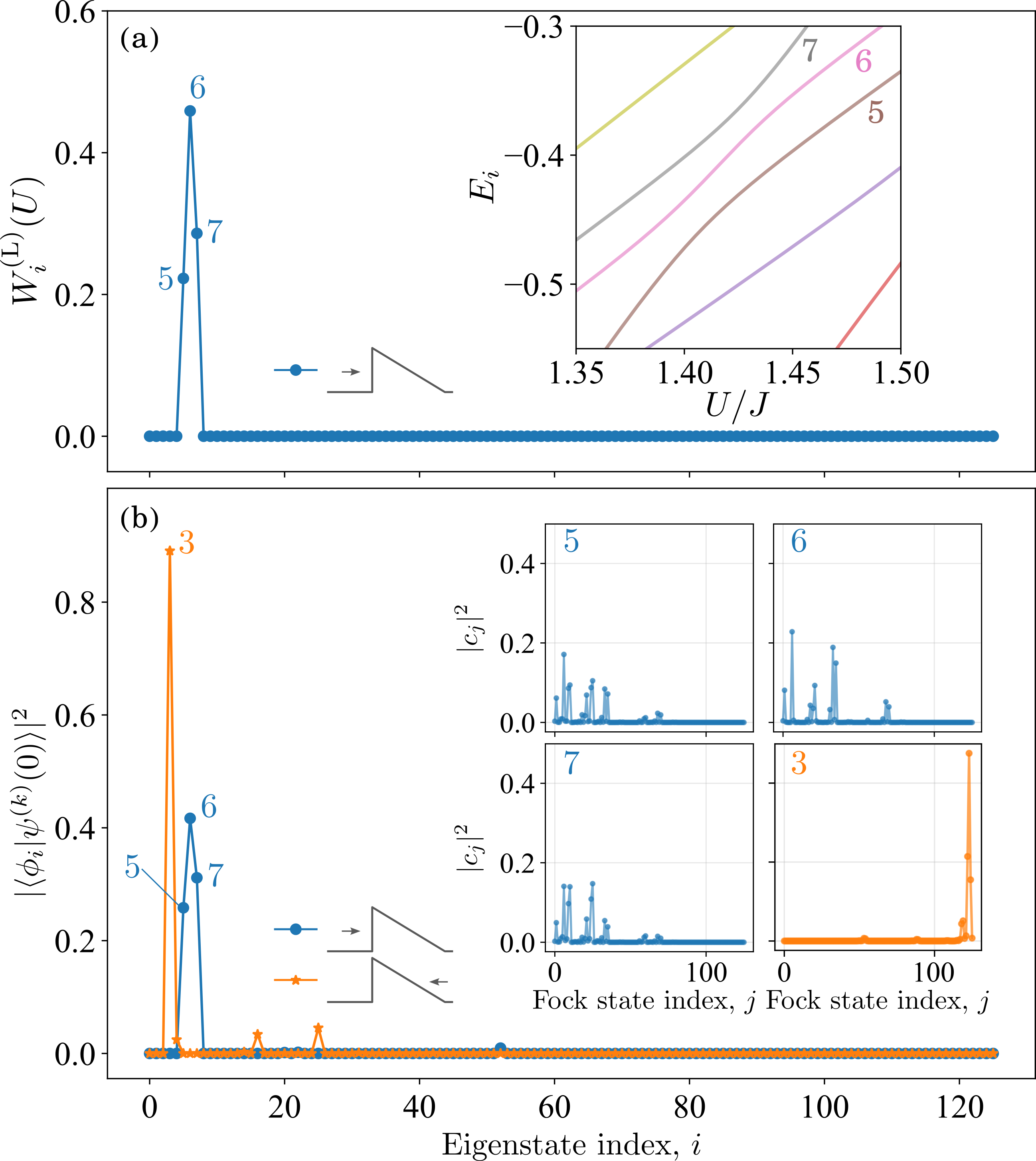}
    \caption{{Eigenstate structure underlying directional transport at $U = 1.42J$ and $h = 10J$ for a system of size $L = 6$ with $N = 4$ bosons. (a)~Individual eigenstate contributions $W^{\rm (L)}_i(U)$ to the long-time average post-barrier population. Inset: eigenenergies $E_i$ vs $U/J$ corresponding to eigenstates contributing most~($i=5, 6, 7$) to $W^{\rm (L)}(U)$. (b)~Overlap $|\langle\phi_i|\psi^{(k)}(0)\rangle|^2$ between the eigenstates of the Hamiltonian $\hat{\mathcal{H}}$ and the initial states with left-localized (blue) and right-localized (orange) bosons. Bosons initially localized from the right project almost entirely onto a single eigenstate ($i=3$, overlap $\approx 0.9$), while bosons initially localized from the left create a superposition of three eigenstates ($i=5,6,7$, maximum overlap $\approx 0.4$). Insets: Fock-state composition $|c_j|^2$ of the dominant eigenstates. The top five Fock-state contributions to these eigenstates are detailed in Table~\ref{tab:eigenstates}.}}
    \label{fig:overlap}
\end{figure}

Crucially, the width of each transport window is governed by the rate $dE_i/dU$ at which the participating energy levels separate. For the narrow peak near $U \approx 2.25J$ [Fig.~\ref{fig:transport_windows}(d)], the energy levels approach closely but diverge rapidly, with the crossing region spanning only $\Delta U \sim 3.6 \times 10^{-4}$. By contrast, the broad transport window at $U \approx 1.42J$ arises because the relevant energy levels run nearly parallel over an extended interval $\Delta U \sim 0.05J$ [see Fig.~\ref{fig:overlap}(a) inset].

{\section{Microscopic Mechanism of Directional Asymmetry}} 

\begin{table}[h]
\centering
\caption{Dominant Fock-state contributions to eigenstates with largest initial state overlap for $N=4$ bosons on $L=6$ sites at $U=1.42J$, $h=10J$. Site notation: $|n_1n_2n_3n_4n_5n_6\rangle$.}
\begin{tabular}{cccc}
\hline
Eigenstate & Fock State & Index $j$ & $|c_j|^2$ \\
\hline
3 (from right) & $|000022\rangle$ & 123 & 0.476 \\
 & $|000031\rangle$ & 122 & 0.214 \\
 & $|000013\rangle$ & 124 & 0.155 \\
 & $|000112\rangle$ & 119 & 0.051 \\
 & $|000121\rangle$ & 118 & 0.044 \\
\hline
5 (from left) & $|120001\rangle$ & 25 & 0.147 \\
 & $|220000\rangle$ & 6 & 0.141 \\
 & $|210001\rangle$ & 10 & 0.140 \\
 & $|120010\rangle$ & 24 & 0.109 \\
 & $|210010\rangle$ & 9 & 0.097 \\
\hline
6 (from left) & $|220000\rangle$ & 6 & 0.228 \\
 & $|110020\rangle$ & 33 & 0.189 \\
 & $|110002\rangle$ & 35 & 0.149 \\
 & $|130000\rangle$ & 21 & 0.093 \\
 & $|310000\rangle$ & 1 & 0.081 \\
\hline
7 (from left) & $|220000\rangle$ & 6 & 0.171 \\
 & $|120001\rangle$ & 25 & 0.105 \\
 & $|210001\rangle$ & 10 & 0.094 \\
 & $|120010\rangle$ & 24 & 0.088 \\
 & $|210010\rangle$ & 9 & 0.086 \\
\hline
\end{tabular}
\label{tab:eigenstates}
\end{table}
To get a deeper understanding of the directional asymmetry, we focus on the interaction strength $U = 1.42J$, corresponding to the broadest transport window for particles initially localized from the left [Fig.~\ref{fig:transport_windows}(a)]. We decompose $W^{(\rm L)}(U)$ into individual eigenstate contributions $W_i^{(\rm L)}(U) = |\langle\phi_i(U)|\psi^{(\rm L)}(0)\rangle|^2 \langle\phi_i(U)|n^{(\rm L)}_\mathrm{after}|\phi_i(U)\rangle$ [see Fig.~\ref{fig:overlap}(a)], each of which is a product of two factors: the overlap of the initial state with eigenstate $|\phi_i\rangle$, and the expectation value of $n_\mathrm{after}^{(\rm L)}$ in that eigenstate. The peak in $W^{(\mathrm{L})}(U)$ is dominated by eigenstates $i = 5, 6, 7$, with all other contributions remaining negligible. As discussed in Sec.~\ref{sec:transport_windows}, the corresponding energy levels run nearly parallel over an extended interval $\Delta U \sim 0.05J$ [Fig.~\ref{fig:overlap}(a) inset], which accounts for the width of this transport window. We now ask why the same interaction strength supports transport for left-localized particles while suppressing it for right-localized particles.

{Figure~\ref{fig:overlap}(b) shows that the two initial states project onto the eigenstates in a strikingly different manner.} For particles initially positioned from the {right} side (orange curve), the dynamics are dominated by a single eigenstate ($i=3$) with overlap approaching 0.9, indicating near-perfect projection onto the single state. In stark contrast, initial localization from the {left} side (blue) exhibits fragmented overlap across multiple eigenstates ($i=5, 6, 7$) with maximum overlap $\approx0.4$ signaling a superposition that drives the observed transport. The insets display the Fock-state composition $|c_j|^2$ of the dominant contributing eigenstates (with the top five Fock-state contributions detailed in Table~\ref{tab:eigenstates}). Crucially, {the eigenstate dominating the right-initialized dynamics} (bottom right) comprises Fock states with zero post-barrier occupation, creating an effective transport barrier. {The eigenstates dominating the left-initialized dynamics} (top panels, bottom left), however, contain substantial weight on configurations with particles distributed across the barrier, enabling tunneling. This eigenstate analysis directly connects the macroscopic directional asymmetry to the microscopic many-body structure: {the asymmetric barrier geometry selects fundamentally different superpositions of Fock states depending on the initial localization}, yielding opposite transport behaviors.

\section{Dependence of ``Black-Hole'' Regime on Particle Number and System Size}
\begin{figure}[h]
\centering
\includegraphics[width=0.99\linewidth]{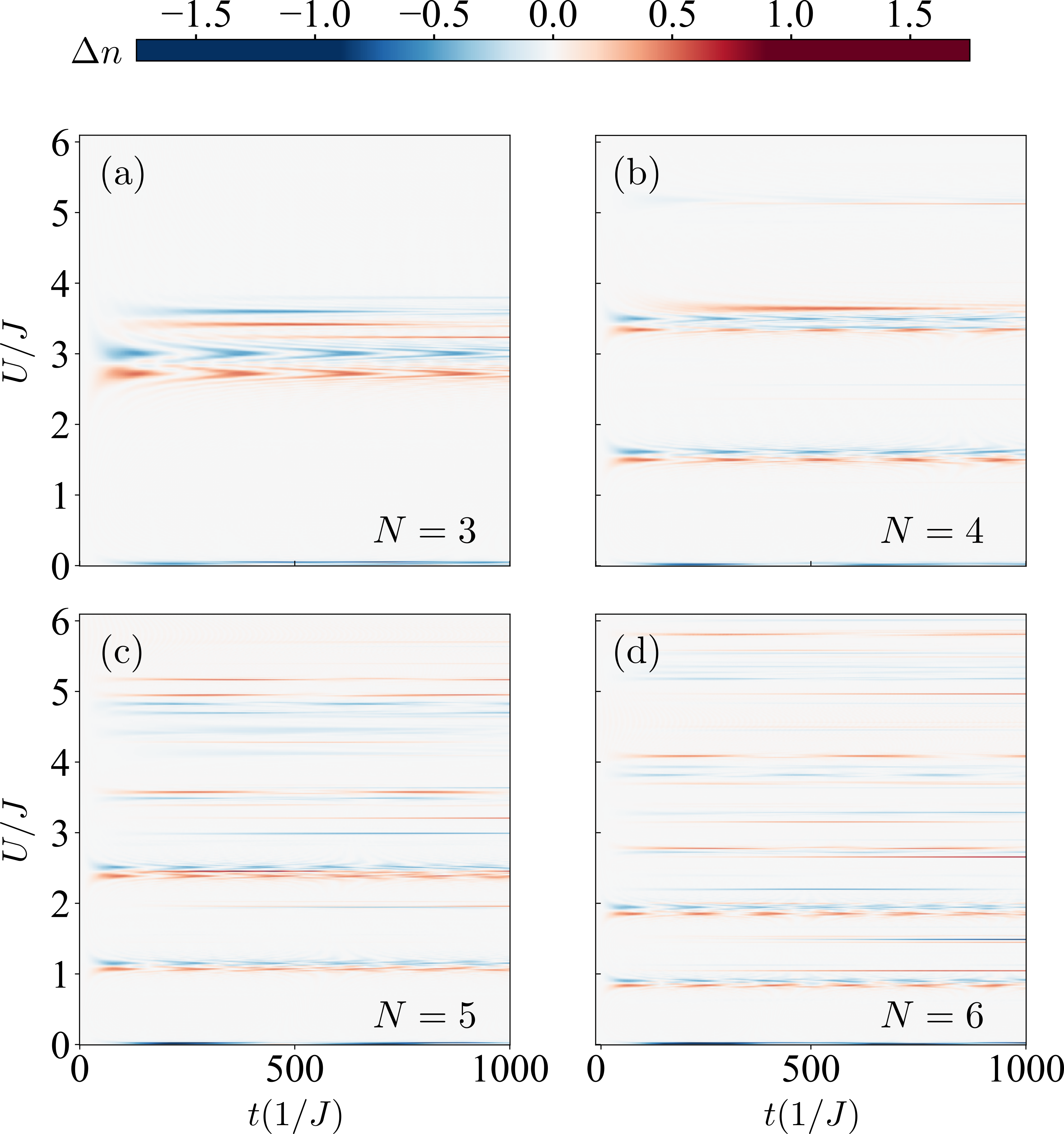}
\caption{Population imbalance $\Delta n$ as a function of time $t$ and interaction strength $U$ for varying particle numbers: (a) $N=3$, (b) $N=4$, (c) $N=5$, and (d) $N=6$. System parameters: $L=8$ sites, barrier height $h=10J$, open boundary conditions. Red (blue) regions indicate preferential tunneling from the {left (right)} side.}
\label{fig:N-dependence}
\end{figure}

In this section, we analyze how the ``black-hole'' region---identified by a large and persistent population imbalance $\Delta n$---depends on the total number of bosons in the system and the system size. We begin by examining the particle number dependence for a fixed chain of length $L=8$ with barrier height $h=10J$. The total particle number $N$ is varied as indicated in Fig.~\ref{fig:N-dependence}, specifically $N=3,4,5,6$.

For $N=3$ particles [see Fig.~\ref{fig:N-dependence}(a)], the system exhibits multiple directional transport regimes within a continuous interaction window $2.5J \lesssim U \lesssim 3.75J$. These regimes alternate between favoring tunneling from {left} side (red regions, positive $\Delta n$) and tunneling from {right} side (blue regions, negative $\Delta n$), with blue bands typically following red ones. As the particle number increases, the structure of the population imbalance undergoes a qualitative change. At $N=4$ [Fig.~\ref{fig:N-dependence}(b)], the single broad region splits into two spatially separated bands centered around $U\approx1.5J$ and $U\approx3.5J$. Increasing the filling further to $N=5$ [Fig.~\ref{fig:N-dependence}(c)] and $N=6$ [Fig.~\ref{fig:N-dependence}(d)] enhances this fragmentation.
\begin{figure}[h]
\centering
\includegraphics[width=\linewidth]{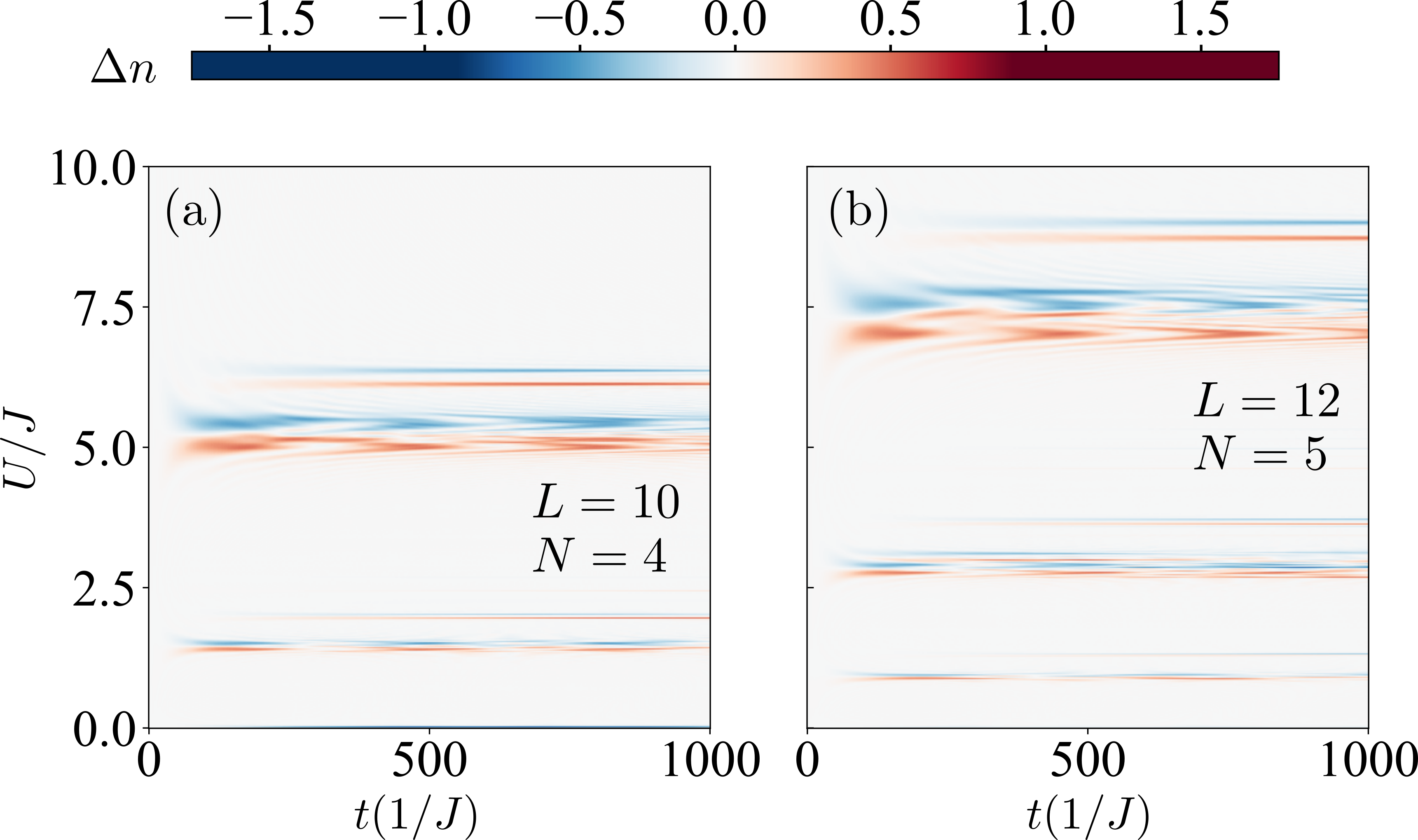}
\caption{Population imbalance $\Delta n$ as a function of time $t$ and interaction strength $U$ for: (a) $L=10$ and $N=4$, (b) $L=12$ and $N=5$; barrier height $h=10J$, open boundary conditions. Red (blue) regions indicate preferential tunneling from the {left (right)} side.}
\label{fig:broad_bands}
\end{figure}
The splitting reflects the restructuring of the many-body spectrum at higher filling. Adding particles introduces interaction-assisted tunneling channels and differentially affects the tunneling dynamics for {two different initial} configurations.

Having established how particle number affects the transport windows at fixed system size, we now examine how these features scale with the lattice length while maintaining low particle density. Figure~\ref{fig:broad_bands} shows results for two representative cases: $L=10$ sites with $N=4$ particles [Fig.~\ref{fig:broad_bands}(a)] and $L=12$ sites with $N=5$ particles [Fig.~\ref{fig:broad_bands}(b)]. {In both cases, sites occupied initially maintain approximately one boson per site}. As before, the barrier height is fixed at $h=10J$. 
Both system sizes exhibit a relatively broad window of large population imbalance. However, with the increase of the system size, the window shifts to higher interaction strengths $U$. Within these broad windows, the characteristic alternating pattern persists, with blue regions typically following red regions. Beyond the primary broad window, both systems display additional narrow stripes of large population imbalance at lower interaction strengths. Notably, the larger system exhibits a greater number of such stripes. The persistence of both broad and narrow directional transport windows across system sizes demonstrates that this phenomenon is not a finite-size artifact but rather a fundamental feature of interaction-driven dynamics in asymmetric barriers.

These results have important experimental implications: while the transport effect persists across all particle numbers studied, experiments at lower filling fractions will observe broader, more experimentally accessible parameter windows for directional tunneling.

\section{Conclusions}
We have demonstrated that interacting bosons in an optical lattice with an asymmetric barrier exhibit {interaction-driven directional} quantum transport without external driving, dissipation, or coupling to reservoirs. This directionality {is a joint action of barrier geometry, many-body interactions, and initial localiozation}, which establishes a new route to transport rectification in closed quantum systems. The observed directionality is present for Fock state as well as for coherent state initial conditions and for a broad range of system parameters.
Eigenstate analysis reveals the mechanism: the asymmetric barrier projects initial states onto manifolds with opposite transport properties. From one direction, the system projects onto eigenstates comprising Fock states with zero post-barrier occupation, forbidding transport. From the opposite direction, it projects onto eigenstates with significant post-barrier occupation, enabling transport. This creates an effectively one-way dynamical boundary in Hilbert space---a quantum analog of a black hole event horizon.
The interaction-driven mechanism persists across varying particle numbers and system sizes {with the relevant interaction window shifting to higher $U$ as system size increases. Experimental precision for the broadest transport window~($\delta U / U = 3.5\%$) is} readily achievable via Feshbach resonance control. {Our results identify interactions as an intrinsic mechanism for directionality, establishing a coherent quantum analog of a diode operating without dissipation and opening new possibilities for atomtronic circuits and coherent many-body transport control.}

\acknowledgments

E.B. was supported by the National Science Foundation (NSF) IMPRESS-U Grant No.~2403609. D.I.B. was supported by Army Research Office (ARO) (grant W911NF-23-1-0288; program manager Dr.~James Joseph). V.B. is supported by NASA EPSCoR and/or the Board of Regents Support Fund. The views and conclusions contained in this document are those of the authors and should not be interpreted as representing the official policies, either expressed or implied, of ARO, NSF, or the U.S. Government. The U.S. Government is authorized to reproduce and distribute reprints for Government purposes notwithstanding any copyright notation herein.

V.B. and E.B. contributed equally to this work.

\bibliography{main}

\end{document}